\documentclass[twocolumn,epjc3]{svjour3}          

\RequirePackage[T1]{fontenc}

\smartqed  

\RequirePackage{graphicx}
\RequirePackage{mathptmx}      
\RequirePackage{flushend}
\RequirePackage[numbers,sort&compress]{natbib}
\RequirePackage[colorlinks,citecolor=blue,urlcolor=blue,linkcolor=blue]{hyperref}

\journalname{Eur. Phys. J. C}

\begin{document}

\title{The golden ratio in Schwarzschild--Kottler black holes}

\author{Norman Cruz\thanksref{e1,addr1}
        \and
        Marco Olivares\thanksref{e2,addr2} 
        \and
        J. R. Villanueva\thanksref{e3,addr3}
}

\thankstext{e1}{e-mail: ncruz@lauca.usach.cl}
\thankstext{e2}{e-mail: marco.olivaresr@mail.udp.cl}
\thankstext{e3}{e-mail: jose.villanuevalob@uv.cl}

\institute{Departamento de F\'{\i}sica, Facultad de Ciencia,
	Universidad de Santiago de Chile, Avenida Ecuador 3493, Estaci\'on Central, Casilla 307, Santiago 2, Chile.\label{addr1}
          \and
          Facultad de Ingenier\'ia, Universidad
          Diego Portales, Avenida Ej\'ercito Libertador 441, Casilla 298--V,
          Santiago, Chile.\label{addr2}
          \and
          Instituto de F\'{\i}sica y Astronom\'ia, 
          Universidad de Valpara\'iso, Avenida Gran Breta\~na 1111, Casilla 5030,
          Valpara\'iso, Chile\label{addr3}
}


\maketitle

\abstract{
In this paper we show that the golden ratio is present in the Schwarzschild--Kottler metric. For null geodesics with maximal radial acceleration, the turning points of the orbits are in the golden ratio $\Phi=(\sqrt{5}-1)/2$. This is a general result which is independent of the value and sign of the cosmological constant $\Lambda$.
}


\section{INTRODUCTION}\label{intro}
The presence of a non--zero vacuum energy (the cosmological constant $\Lambda$) in the main models of theoretical physics such as the superstring and the standard Einstein cosmological models  have motivated consideration of spherical symmetric spacetimes with non-- zero vacuum energy in order to study the well-known effects predicted by General Relativity for planetary orbits and massless  particles in the context of the Schwarzschild spacetime \cite{schwarzschild}, which can be found, for example, in \cite{Chandrasekhar,Adler,Shutz,Gibbons,Cornbleet}, among others.
This study involves determining the geodesic structure of Kottler spacetimes \cite{kotler} and then using a classical test to proof the influence of $\Lambda$. In this sense, the literature dealing with the application of the classical test of general relativity is extensive. To mention a few, the bending of light was examined by Lake \cite{Lake}, who found that the cosmological constant produces no change in this effect; Kraniotis and Whitehouse \cite{K-W}  obtained the compact calculation of the perihelion precession of Mercury by means of genus--2 Siegelsche modular forms. Both tests were applied by Freire et. al. \cite{freire} in the Schwarzschild--Kottler spacetime plus a conical defect, so they obtained that the parameter characterizing such a conical defect is less than $10^{-9}$.  The study of geodesics is also comprehensive.  Some properties of the motion of test particles on Schwarzschild--Kottler spacetimes can be found in \cite{Hledik2}. Timelike geodesics  for positive cosmological constant were investigated in \cite{Jaklitsch}, using only the method of an effective potential in order to found the conditions for the existence of bound orbits. Analysis of the effective  potential  for radial null geodesics in Reissner--Nordstr\"om de Sitter and Kerr de Sitter spacetimes was performed in \cite{Stuchlik}, whereas some properties of the Reissner--Nordstr\"om black hole and naked singularity spacetimes with a non--zero cosmological constant can be found in \cite{Hledik}. Null geodesics in a charged anti--de Sitter spacetime was studied by Villanueva et al. \cite{villanueva13}. Podolsky \cite{podolsky} investigated all possible geodesic motions for extreme  Schwarzschild de Sitter spacetimes. The motion of massive particles in the Kerr and Kerr anti--de Sitter  gravitational fields was investigated in \cite{Kraniotis}, where the geodesic equations are derived by solving the Hamilton-Jacobi partial differential equation. Equatorial circular orbits in the Kerr de Sitter spacetimes was performed by Stuchl\'ik and Slan\'y \cite{Slany}. A study which included null geodesics and timelike geodesics in Schwarzschild anti--de Sitter spacetimes was conducted in \cite{COV}.

The main purpose of this article is to show a general behavior of non--radial null geodesics, common to Schwarzschild, Schwarzschild de Sitter, and Schwarzschil anti--de Sitter spacetimes. This general property does not depend on the value of the cosmological constant and appears in the ratio between the apastron and periastron of  two non--radial photons, which possess the same constant of motion $E$ but their movements are allowed in regions separated by  the effective potential barrier of the equivalent one dimensional problem for the radial coordinate $r$. We have found that this ratio is the {\it golden ratio} $\Phi=(\sqrt{5}-1)/2$.  We have solved explicitly, in terms of the Jacobi elliptic functions, non--radial null  geodesics  in Schwarzschild--anti de Sitter and Schwarzschild--de Sitter spacetimes.

It is well known that $\Phi$ appears quite frequently in biology, where many growth patterns exhibit the Fibonacci numbers in which the next number is the sum of the previous 2 numbers (1, 1, 2, 3, 5, 8, 13, 21, etc.). The Fibonacci sequence is connected with the golden ratio. What is of interest in biology is the existence of systems that can grow and evolve. Nevertheless, in non--equilibrium phase transitions, which appears for example in condensed matter, it is possible to find this number. S. Dammer et al.  \cite{Dammer} investigated the properties of a direct bond percolation process for a complex percolation parameter $p$. They found that for $p=-\Phi$,\, $1+\Phi$, and $2$, the survival probability of a cluster can be computed exactly.

It has been pointed out by M. Livio \cite{Livio} that the golden ratio appears in the physics of black holes.  A well-known result \cite{Davis} is the infinity discontinuity of the specific heat at some values of the angular momentum and the charge of Kerr--Newman black holes. The specific heat changes from negative  to positive for Kerr black holes when the ratio $a=J/M$ satisfies $a \simeq 0.68\, M$. This last value is very close to the value of the golden ratio $\Phi=0.618033...$, but is not exactly the same. In the study of photon geodesics in gravitational fields described by general relativity, the golden ratio has been reported by Coelho et al. \cite{Coelho}. In that work, the circular photon orbits in the Weyl solution describing two Schwarzschild  black holes were considered. It was found that as the distance between the two black holes increases, photon orbits approach one another and merge when $M_{K} = \Phi L$, where $M_{K}$ is the Komar mass of each black hole. In the context of supersymmetry, Hubsch et al. \cite{Hubsch} found that the golden ratio controlled chaos in the dynamics associated with some supersymmetric Lagrangians. Also, $\Phi$ has been reported in higher dimensional black holes \cite{nieto,nieto2}

In this paper we report how the golden ratio appears in the rather simple field of Schwarzschild black holes with a cosmological constant. Their appearance in the geodesic structure of black holes and their association with a general behavior of null particles was quite surprising for us.

Our paper is organized as follows: In Section \ref{NGS}, we derive the geodesic equations of motion for non--radial photons using the variational problem associated with the  corresponding spacetime metric. Using the effective potential related to the equivalent one--dimensional problem for the $r$ coordinate, we found a Newton type law of force, evaluating the points where the maximum acceleration $\ddot{r}$ occurs. Explicit solutions are found for this case in terms of Jacobi integrals. In these solutions the golden ratio is explicitly shown. Finally, in Section \ref{FR} we discuss our results.

    \section{NULL GEODESICS}\label{NGS}
    As a starting point, we will consider the most general metric for a static, spherically symmetric spacetime with a cosmological constant $\Lambda$, which reads
    
    \begin{equation}
    {\rm d}s^{2}=-f(r)\,{\rm d}t^{2}+\frac{{\rm d}r^{2}}{f(r)}+r^{2}({\rm d}\theta^{2}+sin^{2}\theta\,{\rm d}\phi^{2}),\label{1.1}\end{equation} where $f(r)$ is the lapse function  given by
    
    \begin{equation} f(r)=1-\frac{2M}{r}-\frac{\Lambda}{3}r^{2}.\label{1.2}\end{equation}
    
    From this lapse function and depending on the value of the
    cosmological constant, we can study the location of the horizons by analyzing separately the three different configurations separately:
    \begin{enumerate}
    	\item Schwarzschild case $(\Lambda=0)$: As the cosmological constant vanishes, the spacetime allows a unique horizon (the event horizon), which is located at
    	\begin{equation}
    	\label{ehsc}
    	r_{+}=2M.
    	\end{equation}
    	\item Anti-de Sitter case $(\Lambda=-\frac{3}{\ell^{2}}<0)$: when the cosmological constant is negative, the spacetime allows a unique horizon (the event horizon), which must be the real positive solution to the cubic equation
    	\begin{equation} r^{3}+\ell^{2}r-2M\ell^{2}=0, \label{1.3}\end{equation} and its result is \cite{COV}
    	\begin{equation}
    	r_{+}=\sqrt{\frac{4\,\ell^2}{3}}\,\sinh \left[ \frac{1}{3}{\rm arcsinh} \left(\frac{ 3\,\sqrt{3}\,M}{\ell}\right) \right]. \label{1.4}\end{equation}
    	\item de Sitter case $(\Lambda>0)$: When a positive cosmological constant satisfies $\Lambda<1/9M^2$, the spacetime allows two horizons (the event horizon $r_+$ and the cosmological horizon $r_{++}$), which are obtained from the cubic equation \cite{Jaklitsch} \begin{equation} r^{3}-\frac{3}{\Lambda}r+\frac{6M}{\Lambda}=0. \label{1.5}\end{equation} Therefore, by defining $\Theta=\arccos(-3M\sqrt{\Lambda})/3$, their expressions are given by
    	\begin{equation}
    	r_{+}=\frac{1}{\sqrt{\Lambda}}\left(\sqrt{3} \sin \Theta-\cos \Theta \right),\label{1.6}\end{equation}
    	and
    	\begin{equation}
    	r_{++}=\frac{2}{\sqrt{\Lambda}}\cos \Theta.\label{1.7}\end{equation}
    	
    \end{enumerate}
    The geodesic motion of photons in a spacetime described by
    (\ref{1.1})--(\ref{1.2}) can be obtained by solving the Euler-Lagrange equations
    associated with this metric (see \cite{{COV},{Adler},{Chandrasekhar}},
    for instance):
    \begin{equation} \dot{\Pi}_{q} - \frac{\partial
    	\mathcal{L}}{\partial q} = 0,\label{1.9}
    \end{equation} where
    $\Pi_{q} = \partial \mathcal{L}/\partial \dot{q}$ is the generalized
    conjugate  momentum of the coordinate $q$. Recalling that
    for massless particles $(\frac{{\rm d}s}{{\rm d}\tau})^2=2\mathcal{L}=0$, the Lagrangian
    is given by \begin{equation}
    \mathcal{L}=-\frac{1}{2}f(r)\,\dot{t}^{2}+\frac{1}{2}f^{-1}(r)\,\dot{r}^{2}+\frac{1}{2}\,r^{2}\,\dot{\theta}
    ^{2}+\frac{1}{2}\,r^{2}\,\sin^{2}\theta\, \dot\phi^2, \label{1.8}
    \end{equation} where a dot represents the derivative with respect to an affine parameter, $\tau$, along the geodesic. Clearly  ($t,\phi$) are cyclic coordinates, so their corresponding conjugate momenta are conserved giving a place the following expressions
    \begin{equation}\Pi_{t} = -f(r)\, \dot{t} = -\sqrt{E}, \label{1.10}
    \end{equation}
    and
    \begin{equation}\Pi_{\phi} = r^{2}\,\sin^{2}\theta\, \dot{\phi}
    = L, \label{1.11}
    \end{equation} where $E$ and $L$ are constants of
    motion.  Since the metric (\ref{1.1}) is asymptotically flat only when $\Lambda=0$,
    the constant of motion $E$ can be associated with the energy for the Schwarzschild case.
    On the other hand, since the motion is confined to an invariant plane,  without loss of
    generality we can choose $\theta=\pi/2$ so $\dot{\theta}=0$.
    Therefore, using (\ref{1.10}) and (\ref{1.11}) into Eq. (\ref{1.8}),
    we obtain the equation of motion for the unidimensional equivalent problem \begin{equation}\dot r^{2}= E-V_{eff}(r).
    \label{1.15}\end{equation} where $V_{eff}$ defines an
    \textit{effective potential} given by
    \begin{equation}V_{eff}=\frac{L^2\,f(r)}{r^2}.
    \label{1.16}\end{equation} In Fig.1, we plot the effective
    potential as a function of the radial coordinate for the
    Schwarzschild case $\Lambda=0$, the  Schwarzschild anti--de Sitter case $\Lambda<0$, and the Schwarzschild de Sitter case $\Lambda>0$.
    \begin{figure}[h!] \label{F1} \begin{center}
    		\includegraphics[width=85mm]{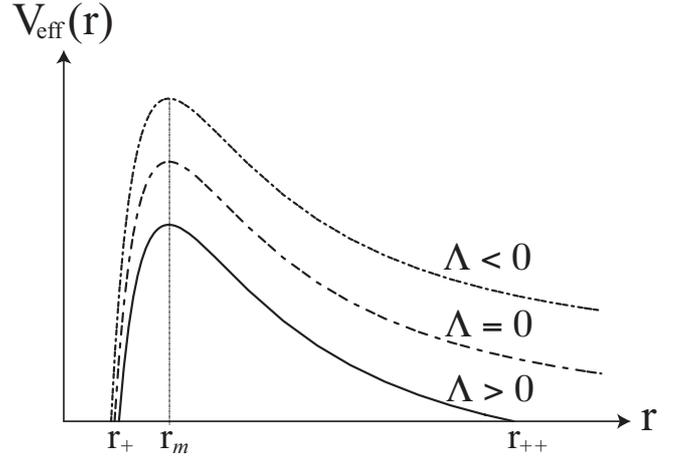}
    		\includegraphics[width=85mm]{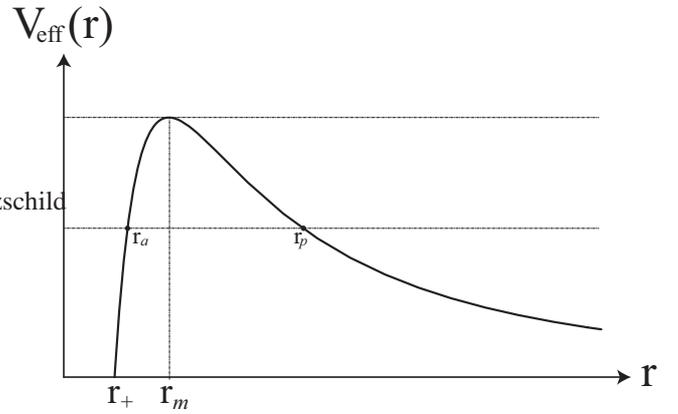}
    		\caption{The figure shows the typical effective potential for non-radial photons in the three cases:
    			$\Lambda=0$ (Schwarzschild case), $\Lambda < 0$ (SAdS case) and $\Lambda > 0$ (SdS case).
    			The maximums are coincident at $r_m=3M$, regardless of the value of the cosmological constant $\Lambda$.}
    	\end{center} \end{figure}

\begin{figure}[!h]
    		\begin{center}
    			\includegraphics[width=85mm]{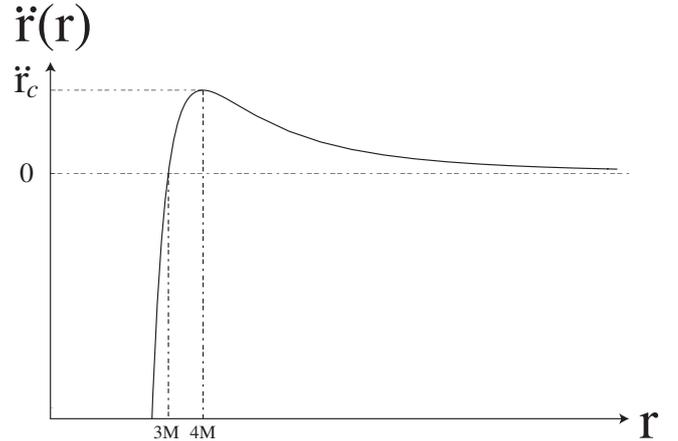}
    		\end{center}
    		\caption{The plot shows the radial acceleration $\ddot{r}$ as a function of the coordinate $r$.
    			It possesses a maximum at $r=4M$ and goes to zero at $r=3M$. So for $r<3M$,
    			the photons possess a negative radial acceleration, which corresponds to the fact that the photons
    			are falling into the event horizon}.
    		\label{f2}
    	\end{figure}

    	Differentiation of the equation (\ref{1.15}) with respect to the
    	affine parameter $\tau$ allows us to find a  Newton type
    	law of effective force for the radial coordinate given
    	by
    	
    	\begin{equation}
    	\ddot r=-\frac{d(V_{eff})}{dr}=\frac{2L^2\,(r-3M)}{r^4}.\label{1.18}
    	\end{equation}
    	This radial acceleration is an indication of the variation
    	of the radial coordinate due to the curvature of the photon
    	trajectory. For radial photons with $L=0$ this acceleration is zero,
    	as we can see from the above equation. Notice that the above
    	expression is independent of the cosmological constant $\Lambda$,
    	which implies that the location of the maximum of the effective
    	potential $r_m=3M$ is common for the three spacetimes (see Fig.1).
    	In other words, the zero effective force on the photons is independent
    	of $\Lambda$. Also, notice that the radial acceleration has a
    	maximum at $r_c=4M$, equal to \begin{equation} \label{acelmax}
    	\ddot{r}_c=\frac{L^{2}}{128\,M^{3}}.
    	\end{equation} In  Fig.\ref{f2} we show the radial acceleration $\ddot{r}$ as a function of the radial coordinate $r$.
    	
    	When the photons possess the maximum radial effective acceleration, their impact parameter $b=L/\sqrt{E}$ becomes
    	\begin{equation}
    	b_{\Phi}=\left(\frac{1}{32M^{2}}-\frac{\Lambda}{3}\right)^{-\frac{1}{2}},
    	\label{1.19}\end{equation} whereas when the photons possess zero radial acceleration, their energies read \begin{equation}b_{0}=\left(\frac{1}{27M^{2}}-\frac{\Lambda}{3}\right)^{-\frac{1}{2}}.
    	\label{1.20}\end{equation} From the two last equations, it is not hard to prove that $b_0<b_{\Phi}$. 
    	
    	Our goal is perform a description of the orbits of the first and
    	second kind, which represent the orbits for photons with $b_0<b<\infty$, so the effective potential imposes the existence of a
    	turning point, $r_a$ for orbits of the first kind, and $r_p$ for
    	orbits of the second kind (see right panel of Fig.1). Therefore, we start considering the zeros
    	of Equation (\ref{1.15}), which obliges us to solve the cubic
    	equation
    	\begin{equation}\mathcal{P}_3(r)\equiv  r^{3}-\mathcal{B}^2\, r+2\,M\,\mathcal{B}^2=0,\label{1.21}\end{equation} where $\mathcal{B}$ is the {\it anomalous impact parameter} defined by the relation \cite{COV} \begin{equation}
    	\label{aip} \frac{1}{\mathcal{B}^2}=\frac{1}{b^2}+\frac{\Lambda}{3}.
    	\end{equation}
    	Notice that in the  Schwarzschild case the anomalous impact parameter coincides
    	with the usual impact parameter. Also, from Eqs. (\ref{1.19}), (\ref{1.20}) and (\ref{aip})
    	it is not hard to see that $\mathcal{B}_0=\sqrt{27}\,M=\sqrt{3}\,r_m$ and $\mathcal{B}_{\Phi}=\sqrt{32}\,M=\sqrt{2}\,r_c$,
    	so, by defining
    	\begin{equation}\Upsilon=\frac{2\sqrt{3}\,\mathcal{B}}{3},
    	\qquad \Xi=\frac{1}{3}\arccos \left(-\frac{\mathcal{B}_0}{\mathcal{B}}\right),\label{1.24}
    	\end{equation}
    	the turning points are given by \begin{eqnarray}
    	\label{turning1} r_{p}&=&\Upsilon \cos\Xi,\\ \label{turning2} r_{a}&=&\frac{\Upsilon}{2} \left(\sqrt{3} \sin \Xi- \cos \Xi \right),
    	\end{eqnarray} whereas the other root of the cubic polynomial (without physical meaning) is given by \begin{equation}
    	\label{negpol}r_n=-\frac{\Upsilon}{2} \left(\sqrt{3} \cos \Xi+ \sin \Xi \right).
    	\end{equation}
    	
    	An important and novel result is found when we consider the ratio between the turning points (\ref{turning1}) and (\ref{turning2}) defined by \begin{equation} \label{ratio1} \zeta(b, M)=\frac{r_a}{r_p}=\frac{\sqrt{3} \tan \Xi- 1 }{2}. \end{equation}Therefore, when massless particles are close to having a maximum radial acceleration, their impact parameter $b\rightarrow b_{\Phi}$, and then we obtain the identity \begin{equation} \label{golden1} \Phi=\lim_{b \rightarrow b_{\Phi}}\zeta(b, M)=
    	{ \sqrt{3}\over 2}\tan\left[\frac{1}{3}\arccos \left(-\sqrt{\frac{27}{32}}\right)\right]-{1 \over 2},\end{equation} where $\Phi=0.618034...=1/(1+\Phi)$ is the golden ratio.
    	An important corollary of the previous statement is obtained in the Scharzschild de Sitter case. From Eq. (\ref{1.19}), $b_{\Phi}\rightarrow \infty$ when $\Lambda=3\,\mathcal{B}_{\Phi}^{-2}$, and therefore, it is not hard to see from Eqs. (\ref{1.6})--(\ref{1.7})  that $r_{++}=4M$ and $r_{+}=4M\Phi$, i.e., the horizons are in the golden ratio.
    	
    	Also, we define the $\xi$ ratio as \begin{equation}
    	\label{ratio2} \xi(b,M)=-\frac{r_n}{r_p}=\frac{\sqrt{3} \tan \Xi+ 1 }{2},
    	\end{equation}
    	and thus $\xi=1+\Phi=1/\Phi$ when $b\rightarrow b_{\Phi}$. Notice that the two last definitions make it possible to write the polynomial (\ref{1.21})  as $\mathcal{P}_3(r)=|r-r_p|\,|r-\zeta\,r_p|\,(r+\xi\,r_p)$, so, using Eqs. (\ref{1.11})--(\ref{1.15}), and then introducing the new variable $u=1/r$, the equation of motion reads
    	\begin{equation}
    	\label{polargenu}\left(-\frac{{\rm d}u}{{\rm d}\phi}\right)^2=2M\,\left|u_p-u \right|\,\left|\frac{u_p}{\zeta}-u \right|\,\left(\frac{u_p}{\xi}+u\right),
    	\end{equation} where $u_p=1/r_p$.

    	\subsection{The golden motion}\label{ofk}
    	As previously mentioned, when the motion of photons is characterized by an impact parameter equal to $b_{\Phi}$, Eqs. (\ref{turning1}) (\ref{turning2}) and (\ref{negpol})  imply that  $r_p=4M$, $r_a=4M\,\Phi$ and $r_n=-4M/\Phi$. Therefore, for orbits of the first kind $r>4M$, and the equation of motion (\ref{polargenu}) becomes \begin{equation}\left(\frac{du}{d\phi}\right)^{2}= 2M\left(\frac{1}{4M}-u\right)\left(u+\frac{\Phi}{4M}\right)\left(\frac{1+\Phi}{4M}-u\right).\label{3.8} \end{equation}
    	Performing the change of variable suggested in \cite{Chandrasekhar,COV},
    	
    	\begin{equation} u=u_{p}\left[1-\frac{\Phi}{2}(1+\cos\chi)\right]\qquad (u=u_{p}\,\,\, \textrm{when}\,\,\, \chi=\pi),\label{3.9}
    	\end{equation}
    	we obtain the following quadrature
    	\begin{equation}\left(\frac{d\chi}{d\phi}\right)^{2}=
    	\frac{2\,\Phi+1}{2}\left(1-k\,\sin^{2}\frac{\chi}{2}\right),\,\qquad \\ \textrm{with}\quad k=\frac{\Phi+1}{2\,\Phi+1}.\label{3.10}
    	\end{equation}
    	Therefore, the solution for the angular coordinate $\phi$ is given by
    	\begin{equation}\phi=\frac{1}{\alpha}\left[K(k)-F\left(\frac{\chi}{2}, k\right)\right],\label{3.12}
    	\end{equation} where $F(\psi,k)$ is the incomplete elliptic integral of the first kind, $K(k)\equiv F(\pi/2,k)$ is the complete elliptic integral of the first kind, and $\alpha=(5/64)^{1/4}$. Therefore, inverting this last equation, and returning to the original variable, we obtain the equation of the orbit of the first kind
    	\begin{equation}r(\phi)=\frac{4M}{1-\Phi\,\textrm{cn}^{2}\left(K(k)-\alpha\,\phi\right)},\label{3.13}
    	\end{equation}where ${\rm cn}(u)\equiv {\rm cn}(u, k)$ is the Jacobi elliptic cosine function.
    	
\begin{figure}[!h]
    		\begin{center}
    			\includegraphics[width=85mm]{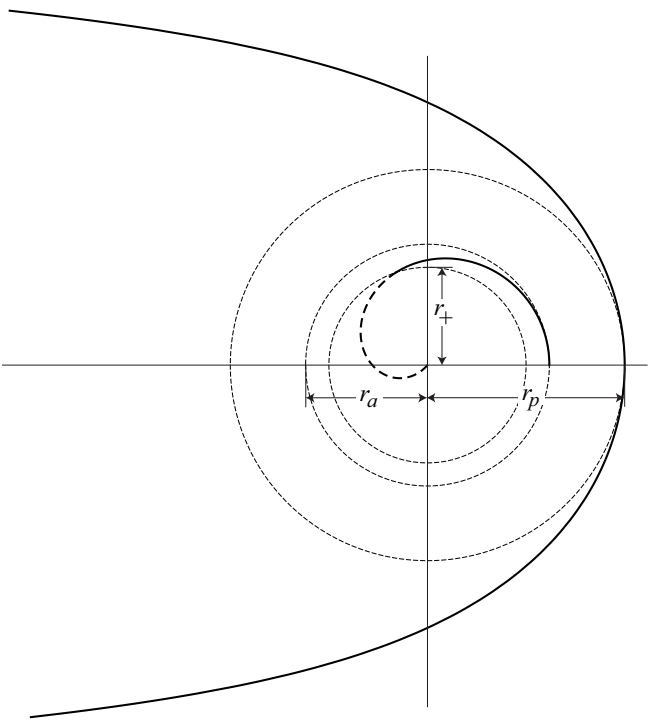}
    		\end{center}
    		\caption{The polar plot shows the null geodesic of the first and second kind. These trajectories correspond to the motion
    			of photon, which possess an impact parameter $b=b_{\Phi}$, such that $r_{p}=4M$ and $r_{a}=\Phi\,r_p$, where $\Phi=0,618034...$ is the golden ratio.}
    		\label{f3}
    	\end{figure}

    	Additionally, for orbits of the second kind we have that $r\leq 4M\Phi$, and the equation of motion (\ref{polargenu}) is given by
    	
    	\begin{equation}\left(\frac{du}{d\phi}\right)^{2}=2M\left(u-\frac{1}{4M}\right)
    	\left(u+\frac{\Phi}{4M}\right)\left(u-\frac{1}{4M\Phi}\right).\label{3.14}\end{equation}
    	In this case, it is possible to obtain an easy quadrature performing the following change of variable:
    	\begin{equation} u=\frac{1}{4M}\left(1+\Phi\sec\frac{\chi}{2}\right),\label{3.15}\end{equation}
    	such that $u=u_{a}$ when $\chi=0$, and $u\rightarrow
    	\infty$ when $\chi\rightarrow \pi$. This substitution reduces Eq. (\ref{3.8}) to the same form as Eq. (\ref{3.10}) with the same value of $k$,  but now it must be written as
    	\begin{equation}\phi=\frac{1}{\alpha}F\left(\frac{\chi}{2}, k\right),\label{3.16}\end{equation} where the zero of $\phi$ is now at the apoastron $r_a=4M\Phi$. Therefore, the trajectory can be obtained by inverting this last equation, resulting in
    	
    	\begin{equation} r(\phi)=\frac{4M}{1+\Phi \,\textrm{nc}(\alpha\,\phi)},\label{3.17}\end{equation}
    	where ${\rm nc}(\psi)=1/{\rm cn}(\psi)$, and ${\rm cn}(\psi)\equiv {\rm cn}(\psi, k)$ is the Jacobi elliptic cosine function. In Fig. \ref{f3} we have plotted the orbits of the first and second kind for photons with impact parameter $b=b_{\Phi}$.
    	
   \section{FINAL REMARKS}\label{FR}
   In this paper we have studied the motion of massless particles in a background described by Schwarzschild--Kottler metric, whose general form is given by Eqs. (\ref{1.1})--(\ref{1.2}). It is given as a solution to the Einstein equations, and is completely determined by its mass $M$ and the cosmological constant $\Lambda$. Here we have presented a review of the spacetime and the corresponding equations of the angular motion, without any restriction on the value of $\Lambda$.
   
   An important feature for this class of spacetime occurs when the acceleration of the radial coordinate is considered. In such a situation, photons with maximum radial acceleration have an impact parameter $b_{\Phi}$, and then their return points are in the golden ratio. This result proves to be independent of the value of the cosmological constant, and allows us to express the golden ratio $\Phi$ as a limit of the function (\ref{golden1}), i.e., $\Phi=\lim_{b \rightarrow b_{\Phi}}\zeta(b, M)$, where $\zeta$ is the ratio between the apoastron and periastron distances. Thus, the golden ratio, which characterizes the fractal structure of nature, also appears  in the geodesic structure of black holes, in particular in the movements of null particles and independently of the value and sign of the cosmological constant $\Lambda$.
   
   The understanding of gravitational fields is strongly linked to geometry: Newton's theory is developed on a three - dimensional plane space in Euclidean geometry. The change that Einstein made was enormous in interpreting spacetime as a curved manifold, i. e., a description of gravity through Riemann's geometry. In this way, when we find the golden ratio in the geodesic structure of black holes, it gives us the future possibility of studying gravitation with fractal geometry, the geometry of nature.

\begin{acknowledgement}
N. C. and M. O.  acknowledges the hospitality of Instituto de F\'isica y Astronom\'ia of Universidad de Valpara\'iso, where part of this work was done. This research was supported by CONICYT through Grant FONDECYT No. 1140238 (NC) and No. 11130695 (JRV)
\end{acknowledgement}


\end{document}